\documentstyle[emulateapj2,onecolfloat,psfig]{article}
\begin{document}
\twocolumn[
\title{The ages of very cool hydrogen--rich white dwarfs}

\author{M. Salaris \altaffilmark{1}, E. Garc\'{\i}a--Berro
\altaffilmark{2,3}, M. Hernanz \altaffilmark{3}, J. Isern \altaffilmark{3}, 
D. Saumon \altaffilmark{4}
}

\affil{$^{1} $Astrophysics Research Institute, Liverpool John Moores 
       University, Twelve Quays House, Egerton Wharf, Birkenhead CH41 
       1LD, United Kingdom}
\affil{$^{2}$ Departament de F\'{\i}sica Aplicada, Universitat Polit\`ecnica 
       de Catalunya, Jordi Girona Salgado s/n, M\`odul B--5, Campus 
       Nord, 08034 Barcelona, Spain}
\affil{$^{3}$ Institut d'Estudis Espacials de Catalunya/UPC, Edifici Nexus, 
       Gran Capit\`a 2--4, 08034 Barcelona, Spain}
\affil{$^{4}$ Department of Physics and Astronomy, Vanderbilt University,
       P.O. Box 1807 Station B, Nashville, TN~37235}


\begin{abstract}

The  evolution of white dwarfs is  essentially  a cooling  process  that
depends  primarily on the energy stored in their degenerate cores and on
the transparency of their envelopes.  In this paper we compute  accurate
cooling  sequences  for   carbon--oxygen   white  dwarfs  with  hydrogen
dominated  atmospheres  for the full  range of masses of  interest.  For
this purpose we use the most accurate available physical inputs for both
the  equation  of  state  and  opacities  of the  envelope  and  for the
thermodynamic  quantities of the degenerate  core.  We also  investigate
the  role of the  latent  heat in the  computed  cooling  sequences.  We
present  separately  cooling  sequences  in which the  effects  of phase
separation of the  carbon--oxygen  binary  mixture upon  crystallization
have been neglected, and the delay  introduced in the cooling times when
this  mechanism is properly  taken into account, in order to compare our
results with other published  cooling  sequences  which do not include a
treatment  of this  phenomenon.  We find that the  cooling  ages of very
cool   white   dwarfs   with  pure   hydrogen   atmospheres   have  been
systematically   underestimated   by   roughly   1.5  Gyr  at   $\log(L/
L_{\sun})=-4.5$  for an  otherwise  typical  $\sim 0.6  M_{\sun}$  white
dwarf, when phase  separation is neglected.  If phase  separation of the
binary  mixture is included then the cooling ages are further  increased
by roughly  10\%.  Cooling  tracks and  cooling  isochrones  in  several
color--magnitude diagrams are presented as well.

\end{abstract}
]

\section{Introduction}

White dwarfs are well studied  objects and the physical  processes  that
control their evolution are relatively  well  understood.  In fact, most
phases of white dwarf  evolution can be succesfully  characterized  as a
cooling process.  That is, white dwarfs slowly radiate at the expense of
the  residual  thermal  energy of their  ions.  The  release  of thermal
energy  lasts for long time  scales  --- of the  order of the age of the
galactic disk  ($\sim10$  Gyr).  While their  detailed  energy budget is
still today somehow  controversial  --- the reason being  basically  the
release of  gravitational  energy  associated to phase  separation  upon
crystallization  (Mochkovitch  1983;  Garc{\'\i}a--Berro  et al 1988a,b;
Isern et al 1997) --- their  mechanical  structures,  which are  largely
supported by the pressure of the gas of degenerate  electrons,  are very
well  modeled  except for the outer  layers.  These  layers  control the
energy  output and their  correct  modeling  is  necessary  to  properly
understand  the evolution of white dwarfs.  This is  especially  true at
very low luminosities  when most of the white dwarf interior has already
crystallized and the cooling process is controlled almost exclusively by
the  behavior  of  the  very  outer  layers.  The  situation  was  quite
unsatisfactory  until  very  recently,  when  new  developements  in the
physics of white dwarf  atmospheres  allowed the calculation of reliable
hydrogen--dominated   white   dwarf   atmospheres   down  to   effective
temperatures as low as 1500~K (Saumon \& Jacobson 1999; Hansen 1999).

The study of very cool white dwarfs bears important  consequences, since
the recent results of the  microlensing  experiments  carried out by the
MACHO team (Alcock et al 1997) yield that perhaps a substantial fraction
of the halo dark matter could be in the form of very cool white  dwarfs.
The search for this elusive  white dwarfs has not been  successful  yet,
although there are evidences that perhaps the observational counterparts
of these white dwarfs could be the stellar objects recently  reported in
the Hubble  Deep Field  (Ibata et al 1999;  M\'endez  \& Minniti  1999).
Most probably one of the reasons for this failure in detecting very cool
white dwarfs was that their colors were  expected to be redder than they
really are (Hansen  1999).  Moreover,  the  luminosity  function of disk
white dwarfs has been repeatedly  used during the last decade to provide
independent  estimates of the age of galactic  disk  (Winget et al 1987;
Garc{\'\i}a--Berro et al 1988b; Hernanz et al 1994).

In view of the recent  progresses  in the physics of the dense  hydrogen
plasma we compute new cooling sequences which will be invaluable for the
study of the structure and age of our Galaxy and its halo.  The paper is
organized as follows:  in \S2 we briefly describe the evolutionary  code
and the adopted physical inputs, in \S3 we discuss in detail the cooling
sequences, we compare them to other cooling sequences  available so far,
and we show their evolution in the  color--magnitude  diagram.  Finally,
our conclusions are drawn in \S4.

\section{The evolutionary code}

The adopted cooling code is the same  evolutionary  code used by Salaris
et  al~(1997)  in which we have  included an accurate  treatment  of the
crystallization  process of the carbon--oxygen (C/O) core, together with
updated input  physics  suitable for  computing  white dwarf  evolution,
which will be described in the  following  paragraphs.  Thus the cooling
sequences  described in the next section have been computed using a full
evolutionary  code  instead  of  the  approximate  procedure  used,  for
instance, in Garc\'\i a--Berro et al (1996).

The  equation of state (EOS) for the C/O binary  mixture in the  gaseous
phase was taken from  Straniero  (1988),  while for the liquid and solid
phases the detailed EOS from  Segretain et  al~(1994)  was used.  As for
the pure H and He regions, we used the  results of Saumon,  Chabrier  \&
Van Horn (1995),  supplemented at the highest  densities by an EOS for H
and He using the  physical  prescriptions  by  Segretain  et  al~(1994).
Crystallization was considered to occur at $\Gamma=180$,  where $\Gamma$
is the usual plasma ion coupling  parameter.  The associated  release of
latent heat was  assumed to be equal to  $0.77~k_{\rm  B}T$ per ion (see
sections 3.1 and 3.2 below).  The additional energy release due to phase
separation  of  the  C/O  mixture  upon   crystallization  was  computed
following closely Isern et al~(1997, 2000).

Neutrino   energy  losses  were  taken  from  Itoh  et   al~(1996).  The
conductive opacities for the liquid and solid phase of Itoh et al~(1983)
and Mitake,  Ichimaru  \&  Itoh~(1984)  were  adopted;  for the range of
temperatures and densities not covered by the previous  results, we used
the conductivities by Hubbard \& Lampe (1969).  OPAL radiative opacities
(Iglesias \& Rogers 1993) with $Z=0$ were used for  $T\geq6000$~K in the
He and H  envelopes.  In the H envelope,  and for the  temperatures  and
densities  not covered by the OPAL  tables, we computed  Rosseland  mean
opacities from the monochromatic opacities of Saumon \& Jacobson~(1999),
after adding the  contribution of hydrogen lines.  The surface  boundary
conditions  needed to integrate  the stellar  structure  ($P$ and $T$ at
$\tau=200$,  where  the  diffusion  approximation  is valid  and one can
safely start to integrate  the full set of stellar  structure  equations
using Rosseland mean  opacities)  were obtained from detailed  non--grey
model atmospheres:  for $T_{\rm  eff}\leq4000$~K  we used the results of
Saumon \& Jacobson~(1999),  whereas for higher temperatures, the results
of  Bergeron,  Wesemael \&  Beauchamp~(1995)  were  adopted.  Bolometric
luminosities  and effective  temperatures of the white dwarf models were
transformed  into $V$  magnitudes  and  colors by using  the  bolometric
corrections and  color--$T_{\rm  eff}$  relations  derived from the same
model  atmospheres.  Superadiabatic   convection  in  the  envelope  was
treated according to the ML2 parametrization of the mixing length theory
(see Bergeron, Wesemael \& Fontaine 1992, and references therein), which
is the same  formalism  used also in the Saumon \&  Jacobson~(1999)  and
Bergeron, Wesemael \& Beauchamp~(1995) computations.

When  computing  the  cooling  sequences,  for each white  dwarf mass an
initial   model  was   converged  at   log$(L/L_{\odot})\approx2.0$   by
considering a C/O core with the chemical  composition profile taken from
the evolutionary  pre--white dwarf computations of Salaris et al~(1997),
together with ``thick'' H and He layers ($M_{\rm H} =10^{-4}M_{\rm WD}$,
$M_{\rm He}=10^{-2}M_{\rm WD}$), and it was evolved down to luminosities
$\log(L/L_{\odot})\approx  -5.5$, when all the model  sequences  already
have large enough ages.

\section{The cooling sequences}

We have computed cooling sequences  neglecting and including the release
of gravitational  energy  associated to phase  separation with the above
described  physical  inputs for white dwarf masses  $M_{\rm  WD}=0.538$,
0.551, 0.606, 0.682, 0.768, 0.867 and 1.0 $M_{\sun}$, which are the core
masses  derived in Salaris et al (1997).  The  cooling  sequences  for a
typical 0.606 $M_{\sun}$  white dwarf for both cases are shown in figure
1.  The adopted  release of latent  heat for the  calculations  shown in
figure 1 is $k_{\rm B}T$ per ion in order to facilitate  the  comparison
with similar  calculations.  The solid line  corresponds  to the case in
which the release of  gravitational  energy due to phase  separation has
been neglected, whereas the dotted line corresponds to the case in which
the  effects of phase  separation  have been fully  taken into  account.
Also  shown  in  figure  1  are  the  cooling  sequences  computed  with
$l=0.77~k_{\rm  B}T$ per ion for the two above mentioned  cases in which
phase  separation  has been included or neglected  (long dashed line and
dashed dotted line, respectively).

\begin{figure}[t]
\centerline{\psfig{file=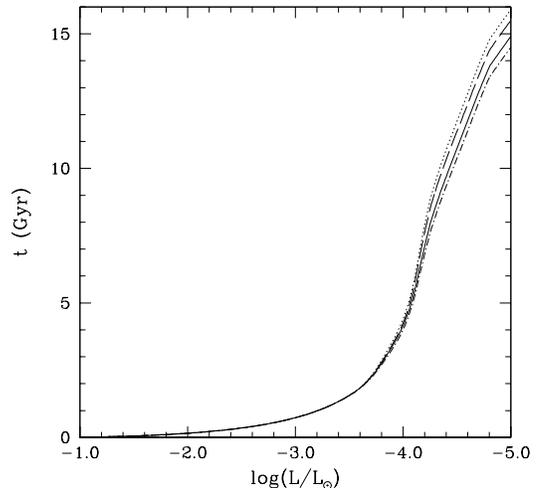,width=2.90in}}
\vskip-0.4cm
\caption{
Cooling  sequences  for a  0.606  $M_{\sun}$  white  dwarf
neglecting phase  separation  (solid line) and taking into account phase
separation (dotted line) computed assuming  $l=k_{\rm B}T$ per particle.
Also shown are the the cooling  sequences  obtained with  $l=0.77~k_{\rm
B}T$ per  particle in both cases  (dashed  dotted  line and long  dashed
line, respectively).
}
\label{fig:fig1}
\end{figure}


\subsection{Comparison with previous evolutionary calculations}

The most distinctive feature of the cooling sequences  presented here is
the duration of the cooling phase itself,  which is larger than in other
equivalent  calculations  (Hansen 1999, Althaus \& Benvenuto 1998).  For
instance,  for the case in which the  effects of phase  separation  have
been neglected and the adopted latent heat is $k_{\rm B}T$ per particle,
at  $\log(L/L_{\sun})=-4.5$  the  calculations  for the 0.606 $M_{\sun}$
white dwarf reported here yield a cooling age of $t\simeq  10.7$ Gyr, in
contrast with the  calculation of Hansen (1999) which, for the same core
mass, the same adopted latent heat and the same adopted envelope, yields
$t\simeq 8.9$ Gyr.  That is, the cooling age of a 0.606 $M_{\sun}$ white
dwarf  derived in the  present  work at this  luminosity  is  $\sim$20\%
larger than the cooling age derived by Hansen (1999).

This difference can be easily accounted for from a detailed  analysis of
figure  2,  where  the  core  temperature--luminosity   ($T_{\rm  c}-L$)
relationship  obtained  in this  work ---  solid  line --- and in Hansen
(1999) --- dotted line --- are shown in the upper panel,  whereas in the
lower panel their relative difference is shown as a function of the core
temperature.  As it can be seen  there,  for a  given  $T_{\rm  c}$  the
luminosity  is about 30\% larger down to  $\log(L/L_{\sun})\simeq  -4.5$
and, thus, the model envelopes of Hansen (1999) are systematically  more
transparent  than our envelopes for the same $T_{\rm c}$, resulting in a
more efficient  cooling of the white dwarf interior.  To be precise, let
us quantify how this  affects the cooling  sequences.  For  luminosities
smaller than $L_0=10^{-2}L_{\sun}$ the contribution of thermal neutrinos
and nuclear  reactions are  negligible  and, thus, one can safely assume
that the sole  contribution  to the  cooling  process is the  release of
binding energy.  Therefore we can write:

\begin{equation}
L=-\frac{dB}{dt}.
\end{equation}

Accordingly,  the  difference in the cooling times  between both cooling
sequences can be easily estimated:

\begin{equation}
\Delta t(L)=\int_{L_0}^{L(T_{\rm c})}
\Big(\frac{1}{L_{\rm BH}}-\frac{1}{L_{\rm TW}}\Big)
\frac{dB}{dT_{\rm c}}dT_{\rm c}
\end{equation}

\noindent  where $L_{\rm  TW}(T_{\rm  c})$ stands for the $T_{\rm  c}-L$
relationship  derived in this paper and $L_{\rm  BH}(T_{\rm  c})$ is the
one derived by Hansen (1999).  We have  independently  computed a set of
binding  energies  with the same  equation  of  state  described  in the
previous  section and used  equation  (2) to obtain an  estimate  of the
difference  introduced by the  differences  in the  transparency  of the
envelope.  At  $\log(L/L_{\sun})=-4.5$  we got $\Delta t\simeq 2.1$ Gyr,
which is in good agreement with the value derived from the  evolutionary
code (1.8 Gyr).  Thus, this  difference  can be mainly  ascribed  to the
differences in the transparency of the adopted model envelopes.

\begin{figure}[t]
\centerline{\psfig{file=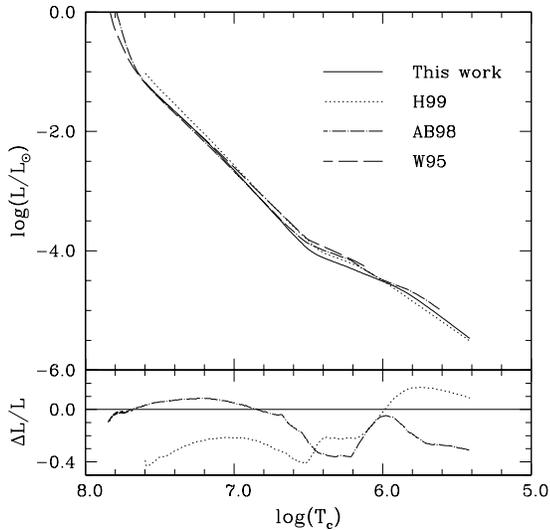,width=3.0in}}
\vskip-0.4cm
\caption{$T_{\rm  c}-L$  relationship for the three cooling sequences
of a $\sim 0.6$  $M_{\sun}$  white dwarf  discussed in this paper (upper
panel).  The solid line corresponds to the relationship  derived in this
work, the dotted line is the relationship  derived by Hansen (1999), and
the dashed--dotted  line is that of Althaus \& Benvenuto (1998).  In the
lower panel the relative  differences  of the latter  cooling  sequences
with our model are shown.  For the sake of completeness we also show the
$T_{\rm  c}-L$  relationship  obtained  by Wood  (1995) as a long dashed
line.
}
\label{fig:fig2}
\end{figure}


Now the question is, why is there a difference  in the model  envelopes?
This question  cannot be answered  categorically.  In fact, we are using
the same  thicknesses for the H and He layers, the same envelope EOS and
the same OPAL  ($Z=0$)  opacities  for  $T\geq6000$~K,  but not the same
model  atmospheres  for the  boundary  conditions.  However  we do  have
indirect ways of checking the consistency of our results.

We recomputed the cooling sequence of our  0.606~$M_{\sun}$  white dwarf
using a grey  $T(\tau)$  relation for deriving the boundary  conditions.
We found that the  $T_{\rm  c}-L$  relationship  derived in this way was
coincident, as long as $T_{\rm  eff}\geq  6000$~K, with the one obtained
using the model atmospheres  boundary  conditions, in agreement with the
findings  by Hansen  (1999).  This  allows us to  directly  compare  our
$T_{\rm  c}-L$  relationship  (for $T_{\rm  eff}\geq  6000$~K)  with the
independent  results of Althaus  \&  Benvenuto  (1998)  for their  0.600
$M_{\sun}$ white dwarf cooling track.  This calculation  adopts our same
metallicity and thickness for the He and H layers.  Moreover, Althaus \&
Benvenuto  (1998)  also used the same EOS in the  envelope  and the same
OPAL  opacities  for  $T\geq  6000$~K.  The  only  differences  of  this
calculation  with respect to our  calculation  and that of Hansen (1999)
are threefold.  First, Althaus \& Benvenuto (1998) used a grey $T(\tau)$
relation for deriving the  boundary  conditions.  As we have shown, this
procedure  is well  justified  as  long  as  $T_{\rm  eff}\geq  6000$~K.
Second,  Althaus \& Benvenuto  (1998)  adopted a slightly  different C/O
profile for the core.  Since the  adopted  internal  C/O  stratification
does not  affect at all the  derived  $T_{\rm  c}-L$  relationship,  the
slightly  different C/O profile  adopted by Althaus \& Benvenuto  (1998)
does not influence the result of the  comparison.  Third, they  employed
the full--spectrum turbulence theory of convection by Canuto, Goldman \&
Mazzitelli   (1996)  instead  of  the  mixing  length  theory,  for  the
computation of the convective  superadiabatic  gradient in the envelope.
However, the treatment of superadiabatic  convection does not affect the
$T_{\rm c}-L$ relationship.

In  the   upper   panel   of   figure   2  we   also   show   the   core
temperature--luminosity  relationship  derived by Althaus  \&  Benvenuto
(1998) as a  dashed--dotted  line,  whereas  in the lower  panel of this
figure the relative  difference with respect to the present  calculation
is shown.  Our result  closely  follows  that of  Althaus  \&  Benvenuto
(1998) in the core temperature range when $T_{\rm eff}\geq  6000$~K, and
therefore it is quite apparent from this figure that for some reason the
model  envelopes  of  Hansen  (1999)  appear  to be, at  least  in  this
temperature  range,  far  more  transparent  than  ours.  It is  however
remarkable that the calculation reported in the present work and that of
Hansen  (1999) are  parallel  for almost the full range of  luminosities
studied here.  Finally both our  calculation  and that of Hansen  (1999)
differ  considerably at low $T_{\rm c}$ from the  calculation of Althaus
\& Benvenuto (1998), as is expected because of the improved  atmospheric
treatment  at very low  luminosities.  For the sake of  completeness  we
also show the  relationship  obtained  by Wood  (1995) as a long  dashed
line (he also used a grey $T(\tau)$ relationship to derive the boundary 
conditions), although we refrain  from doing a detailed  comparison with our
results because this cooling sequence was computed using a different EOS
for the  white  dwarf  envelope.  Note,  however,  that at high  central
temperatures, where different treatments of non-ideal effects should not
substantially affect the envelope EOS, the $T_{\rm c}-L$ relationship of
Wood (1995) is in good agreement with our results.

The  ultimate  reasons for the  differences  found in the  $T_{\rm c}-L$  
relationship (and cooling times) with  respect to  Hansen (1999) results
could  be  various.  First,  it may  have to do, at  least  for  $T_{\rm
eff}\leq 6000$~K (which roughly  corresponds to  $\log(L/L_{\sun})=-3.7$
and  $\log(T_{\rm  c})=6.6$  for our  models)  with the low  temperature
Rosseland mean  opacities  used for  integrating  the stellar  structure
equations, which are not the same in both calculations.

A  contribution  can  arise  also  from the  treatment  of the  boundary
condition  --- see the  discussion  in \S2.1 of Hansen  (1998).  We have
compared the values of $T$ and $P$ at $\tau$=2/3 for  $\log(g)=8$  using
figure 3 of Hansen (1999), and we have found that for a given  effective
temperature  the  values  of the  pressure  used by  Hansen  (1999)  are
systematically higher than ours when $T_{\rm eff}$ is lower than 5500~K.
The  maximum  difference  (in  $\log  P$)  is  $\sim  0.62$  at  $T_{\rm
eff}=4000$~K (which roughly correponds to  $\log(L/L_{\sun})=-4.5$)  and
steadily  decreases  to $\sim  0.41$ at $T_{\rm  eff}=3000$~K  and $\sim
0.18$ at $T_{\rm eff}=2000$~K.  This means that the model atmospheres of
Hansen (1998, 1999) are more  transparent  than ours.  More  transparent
atmospheres  means faster cooling which may explain the  difference.  We
have thus  performed a numerical  experiment  to test the  influence  of
varying the boundary  conditions.  In the hypothesis  that the mentioned
differences of pressure at $\tau$=2/3 exist also at higher values of the
optical depth, we have modified  accordingly our boundary conditions and
repeated  the  computation  of the $\sim 0.61$  $M_{\sun}$  white  dwarf
(whose surface gravity is  $\log(g)=8$).  We have found that the cooling
times  at  luminosities   lower  than   $\log(L/L_{\sun})\sim-4.4$   are
significantly  shorter (up to $\sim$ 1.5 Gyr).  However, at luminosities
lower than about  $\log(L/L_{\sun})=-4.5$  Hansen  (1999)  $T_{\rm c}-L$
relationship  crosses ours, and at a fixed $T_{\rm c}$ the difference in
luminosity  is only of about  10\%  with  respect  to our  results  (our
luminosities   being   higher  in  this   case).  This  shows  that  the
combination of different Rosseland mean opacities and different boundary
conditions  have  a  compensating  effect,  producing  a  $T_{\rm  c}-L$
relationship similar to ours for very dim white dwarfs.

Finally,  we  have  investigated  yet  another  possible  source  of the
difference  between the models of Hansen  (1999) and our models, that is
the  treatment  of the  conductive  opacity  in the H and  He  envelope.
Although we both use for the conductive  opacities in the liquid phase a
combination  of the  Itoh et al  (1983)  and  Hubbard  \&  Lampe  (1969)
results,  we have found  that our  respective  treatments  are  slightly
different  (Hansen,  private  communication;  see also Hansen \& Phinney
1998).  More in detail,  we use Itoh et al (1983)  conductive  opacities
for the liquid phase whenever $ 2 \le \Gamma \le \Gamma_{\rm crist}$ and
$0.0001 \le r_{\rm s} \le 0.5$, where  $r_{\rm s}$ is defined as in Itoh
et al (1983).  Additionally, for the parameter $y$ defined in Itoh et al
(1983)  we  relax  the  constraint  $y<0.1$  discussed  by the  authors,
following  the  recommendations  by Itoh  (1994).  At  lower  densities,
Hubbard \& Lampe (1969) tables are used.  The first two conditions yield
a  lower  limit  for  the  density  which  depends  on the  temperature.
Instead, in the models of Hansen  (1999),  when this limit is lower than
$\log(\rho)\leq  2$, the  tables of  Hubbard  \& Lampe  (1969)  are used
starting  from  $\log(\rho)=2$.  As an example,  for  typical  values of
$\log  P$ and  $\log  T$ at the  base  of the  convective  envelope  for
$\log(L/L_{\sun})\approx-4.5$ we are employing the results of Itoh et al
(1983)  whereas in Hansen  (1998,  1999) the results of Hubbard \& Lampe
(1969)   were  used.  However,   in  the   region   with   $\rho\le10^2$
g~cm$^{-3}$,  where we are still using the Itoh et al (1983)  opacities,
the  Hubbard  \&  Lampe  (1969)  opacities  are  different  by  at  most
$\sim20$\%,  which does not influence  appreciably  the evolution of the
white dwarf.

\subsection{The role of the latent heat}

One of the most  important  sources of energy  during the  evolution  of
cooling white dwarfs is the release of latent heat upon crystallization.
In fact this  source of energy,  together  with the energy  released  by
chemical fractionation, completely dominates the evolution at relatively
low  luminosities  during a  considerable  fraction  of time.  The exact
value of the latent heat is obtained from the  thermodynamic  properties
of  the  plasma   during  the  phase   transition.  In  fact,  both  the
temperature of  solidification  and the precise value of the latent heat
are obtained from the following set of equations:

\begin{eqnarray}
P_{\rm L}(\rho,T)&=&P_{\rm S}(\rho,T)\nonumber\\
&&\\
\mu_{\rm L}(\rho,T)&=&\mu_{\rm S}(\rho,T)\nonumber
\end{eqnarray}

\noindent  where  $T$ is the  crystallization  temperature,  $P$  is the
pressure,  $\mu$ is the chemical  potential, and the  subscripts S and L
stand for the  solid  and the  liquid  phase,  respectively.  The  phase
transition  is solved by providing  the density in the liquid  phase and
solving  equation (3) for the  crystallization  temperature  and for the
density in the solid phase.  As a  consequence,  the solid usually has a
different  density and the latent heat is then the difference in entropy
between both phases.  It has been generally assumed that the latent heat
is of the order of $k_{\rm B}T$ per particle  but, as first  pointed out
by Lamb \& Van Horn  (1975),  the  latent  heat can differ by up to 25\%
from this fiducial  value, the exact value  depending on the density and
temperature of the crystallizing layer.

\begin{deluxetable}{cccccccc}
\footnotesize
\tablecaption{Top section: cooling times, in Gyr, for white dwarfs
of different masses. Bottom section: delay introduced by phase
separation upon crystallization for the same white dwarf models,
also in Gyr.
\label{table1}}
\tablehead{
\colhead{$-\log(L/L_{\sun}$)} & & & &$t$ (Gyr)& & \\
\cline{2-8} &
\colhead{ 0.54 $M_{\sun}$ } &
\colhead{ 0.55 $M_{\sun}$ } &
\colhead{ 0.61 $M_{\sun}$ } &
\colhead{ 0.68 $M_{\sun}$ } &
\colhead{ 0.77 $M_{\sun}$ } &
\colhead{ 0.87 $M_{\sun}$ } &
\colhead{ 1.00 $M_{\sun}$ }
}
\tablewidth{0 pt}
\startdata
2.0 &  0.15 &  0.15 &  0.16 &  0.17 &  0.18 &  0.20 &  0.24 \nl
3.0 &  0.68 &  0.69 &  0.74 &  0.79 &  0.87 &  0.99 &  1.33 \nl
4.0 &  3.56 &  3.62 &  4.00 &  4.34 &  4.81 &  5.17 &  5.28 \nl
4.2 &  6.25 &  6.39 &  6.83 &  7.05 &  7.15 &  6.92 &  6.63 \nl
4.4 &  8.39 &  8.58 &  9.34 &  9.91 & 10.45 & 10.35 &  9.11 \nl
4.6 & 10.26 & 10.50 & 11.40 & 12.06 & 12.67 & 12.91 & 11.80 \nl
4.8 & 12.25 & 12.51 & 13.37 & 13.95 & 14.38 & 14.05 & 12.95 \nl
5.0 & 13.59 & 13.84 & 14.50 & 14.94 & 15.20 & 14.68 & 13.43 \nl
\tableline\\
$-\log(L/L_{\sun}$)& & & &$\delta t$ (Gyr)& & & \\
\cline{2-8}
  & 0.54 $M_{\sun}$ & 0.55 $M_{\sun}$ & 0.61 $M_{\sun}$
  & 0.68 $M_{\sun}$ & 0.77 $M_{\sun}$ & 0.87 $M_{\sun}$
  & 1.00 $M_{\sun}$ \nl
\tableline\\
3.0 &  0.00 &  0.00 &  0.00 &  0.00 &  0.00 &  0.00 &  0.18 \nl
4.0 &  0.12 &  0.15 &  0.25 &  0.48 &  0.82 &  1.00 &  0.98 \nl
4.2 &  0.43 &  0.56 &  0.80 &  1.04 &  1.16 &  1.15 &  1.02 \nl
4.4 &  0.65 &  0.78 &  0.99 &  1.19 &  1.29 &  1.21 &  1.03 \nl
4.6 &  0.67 &  0.80 &  1.00 &  1.19 &  1.29 &  1.21 &  1.03 \nl
\enddata
\nl
\end{deluxetable}

Since most of the computed cooling  sequences adopt this fiducial value,
in the previous  subsection  we have used $k_{\rm B}T$ per particle when
comparing our models with previously  published  results.  It is however
important  to realize  that the latent heat  depends  critically  on the
adopted   equation  of  state  in  both  the  liquid  and  solid  phase.
Therefore, in order to be consistent,  one should compute the release of
latent heat using equation (3) and {\sl the same prescriptions}  adopted
for  the   thermodynamic   quantities  used  to  describe  both  phases.
Moreover,  this  fiducial  value was  derived by Lamb \& Van Horn (1975)
using the best  available  physical  inputs at that  time.  Since  then,
there have been several revisions of the equation of state of very dense
plasmas, being the most significant ones those of Stringfellow,  De Witt
\&  Slattery  (1990) and  Iyetomi,  Ogata \&  Ichimaru  (1993).  We have
solved equation (3) using the chemical  potential and pressures given by
Stringfellow  et al (1990) and Iyetomi et al (1993) and we have obtained
for the range of densities  and  temperatures  relevant  for white dwarf
cooling an almost constant value of $l=0.77~k_{\rm  B}T$ per particle in
both  cases,  which is  considerably  smaller  than the  fiducial  value
adopted  in most  calculations.  This is thus the value we adopt for the
cooling  sequences  described  below.  The net result is,  obviously,  a
decrease in the  cooling  times (see  figure 1) which for a $\sim  0.61\
M_{\sun}$  white  dwarf is  almost  0.4 Gyr at  $\log(L/L_{\sun})=-4.5$.

\subsection{The effects of phase separation}

\begin{figure}[t]
\centerline{\psfig{file=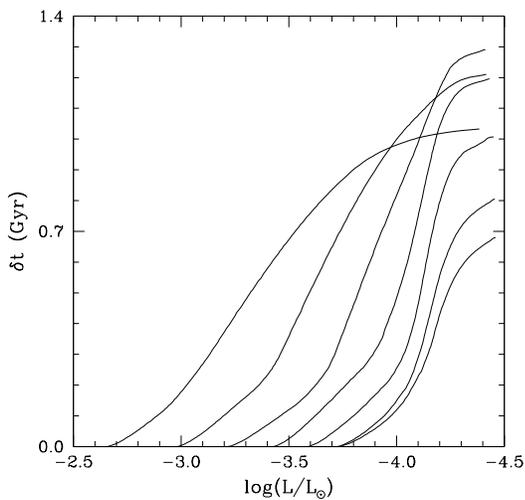,width=2.90in}}
\vskip-0.4cm
\caption{
Computed  delays  due to the C/O  phase  separation  for the
white  dwarf core  masses  discussed  in the text.  From  right to left:
$M_{\rm   WD}=0.538$,   0.551,  0.606,  0.682,   0.768,  0.867  and  1.0
$M_{\sun}$.
}
\label{fig:fig3}
\end{figure}


Now we turn  our  attention  to the  effect  of  phase  separation  upon
crystallization of the C/O binary mixture.  As already noted in Isern et
al (2000), this effect is  differential.  That is, it depends  crucially
not only on the total amount of  gravitational  energy released by phase
separation  upon  crystallization  but also on the  transparency  of the
envelope.   The    delay    introduced    by   phase    separation    at
$\log(L/L_{\sun})=-4.5$  for our $\sim 0.6$ $M_{\sun}$ model is $\sim 1$
Gyr.  Thus the differential effect of phase separation amounts to 9.6\%,
in good agreement with the results of Isern et al (2000).


For the sake of completeness, in figure 3 the delays introduced by phase
separation upon  crystallization are shown for the full range of masses.
None of the  delays is larger  than 1.4 Gyr at  $\log(L/L_{\sun})=-4.5$,
where the  crystallization  process is complete for all but the lightest
white  dwarfs.  It should  be noted as well  that,  for a fixed  $T_{\rm
c}-L$  relationship,  the delays depend on the profile of the C/O binary
mixture,  which varies  considerably  from one core mass to another, and
which is  extremely  dependent  on the  adopted  cross  section  for the
$^{12}$C$(\alpha,\gamma)^{16}$O   reaction  and  on  the  criterion  for
convective instability (Salaris et al 1997) used in the pre--white dwarf
evolutionary  phase.  The results of Salaris et al (1997)  indicate that
the lowest  central  chemical  abundance of $^{16}$O  occurs for a $1.0$
$M_{\sun}$  white dwarf, thus enhancing the effect of phase  separation.
Moreover,  the  size of the  central  oxygen--rich  region  reaches  its
maximum for this mass; this increases the size of the mixing region and,
consequently also increases the effect of phase separation.  Exactly the
opposite occurs for a $0.538$  $M_{\sun}$ white dwarf.  For intermediate
core  masses the  situation  is more  complex  since the two  previously
mentioned  effects are not linear and,  therefore, the analysis of their
combination is more subtle, leading, in any case, to the distribution of
delays shown in figure 3.  However, it is important to realize  that the
delay is  maximum  for the  central  range of white  dwarf  core  masses
(around  $\sim 0.768$  $M_{\sun}$).  In table 1 a summary of the cooling
ages for all the white dwarf masses when phase  separation  is neglected
can be  found\footnote{Detailed  cooling  sequences are  available  upon
request to the authors.}  (upper section), together with the accumulated
time delays  introduced  by chemical  fractionation  at  crystallization
(bottom section).

\begin{table}
\caption{Total gravitational energy released by chemical
fractionation upon crystallization $(E_{\rm g})$ compared to
the total latent heat release $(E_{\rm l})$, both in ergs.
}
\label{tab_mi}
\begin{tabular}{ccc}
\noalign{\smallskip}\hline\noalign{\smallskip}
$M/M_{\sun}$ & $E_{\rm l}$ & $E_{\rm g}$ \\
\noalign{\smallskip}\hline\noalign{\smallskip}
0.54 & $1.22\times 10^{46}$ & $6.27\times 10^{45}$ \\
0.55 & $1.26\times 10^{46}$ & $7.65\times 10^{45}$ \\
0.61 & $1.52\times 10^{46}$ & $1.11\times 10^{46}$ \\
0.68 & $1.92\times 10^{46}$ & $1.66\times 10^{46}$ \\
0.77 & $2.49\times 10^{46}$ & $2.44\times 10^{46}$ \\
0.87 & $3.34\times 10^{46}$ & $3.98\times 10^{46}$ \\
1.00 & $4.90\times 10^{46}$ & $7.37\times 10^{46}$ \\
\noalign{\smallskip}\hline\noalign{\smallskip}
\end{tabular}
\end{table}



Finally, in table 2 we show the total  gravitational  energy released by
phase separation  during the  crystallization  process and we compare it
with the total  release of latent heat.  For the four cooling  sequences
corresponding to the low mass regime ($M_{\rm WD}\sim  0.54,~0.55,~0.61$
and $0.68~M_{\sun}$), the release of gravitational energy (which depends
mostly on the adopted  phase  diagram for the C/O binary  mixture and on
the  pre--white  dwarf  chemical  profile of the C/O mixture) is smaller
than the total release of latent heat --- but it is of the same order of
magnitude --- being almost  identical for $M_{\rm  WD}\sim0.77~M_{\sun}$
white  dwarf.  On the  contrary,  for  the  more  massive  white  dwarfs
(cooling sequences with masses $M_{\rm WD}\sim 0.87$ and 1.0 $M_{\sun}$)
the  total  release  of  latent  heat is  smaller  than the  release  of
gravitational  energy.  Consequently,  we stress  that  this  additional
source  of  energy  cannot  be  neglected   whatsoever  in  a  realistic
calculation.

At  this  point  it  should  be  recalled  that  the  pre--white   dwarf
stratification  of the  C/O  binary  mixture  plays  a  crucial  role in
determining  the  delays.  If a high  effective  rate of the  $^{12}{\rm
C}(\alpha,\gamma)^{16}{\rm  O}$  reaction is adopted  --- as was done in
Salaris et al (1997) --- the  abundance of oxygen in the central  layers
is as high as 0.74 by mass for a typical  0.606  $M_{\sun}$  white dwarf
and the degree of mixing in the liquid layers is strongly reduced by the
very steep gradients of chemical composition, thus minimizing the effect
of  phase  separation.  However,  the  effective  cross  section  of the
$^{12}$C$(\alpha,\gamma)^{16}$O  reaction  rate is the subject of active
debate  and,  consequently,  the  delays  shown in  figure 3  should  be
regarded as a conservative lower limit.

\begin{figure}[t]
\centerline{\psfig{file=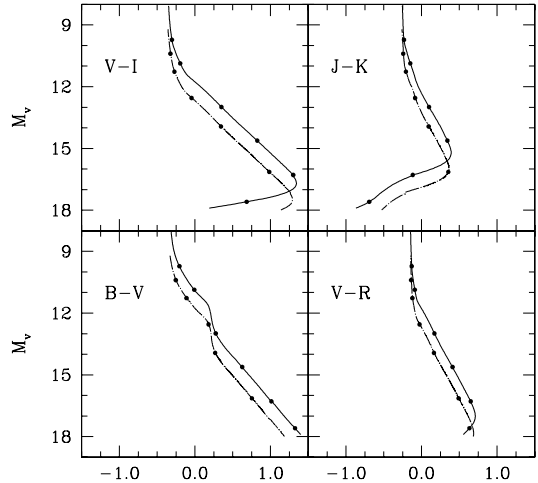,width=3.05in}}
\vskip-0.4cm
\caption{
Evolution  of a 0.606 $M_{\sun}$  white dwarf --- solid line
--- and a 1.0  $M_{\sun}$  white  dwarf ---  dashed--dotted  line --- in
several color--magnitude diagrams.  See text for details.
}
\label{fig:fig4}
\end{figure}


\subsection{The colors of very cool DA white dwarfs}

In figure 4 the  color--magnitude  diagrams for a 0.606 $M_{\sun}$ white
dwarf are shown for four standard  colors.  Starting from the upper left
corner and continuing  clockwise the following  color indices are shown:
$V-I$,  $J-K$, $V-R$ and $B-V$.  The solid line  corresponds  to a 0.606
$M_{\sun}$ white dwarf and the dashed--dotted  line corresponds to a 1.0
$M_{\sun}$ white dwarf.  Selected  evolutionary stages for $\log t=7.0$,
8.0, 9.0, 9.5, 10.0 and 10.2 are represented as dots.  Both evolutionary
tracks are for the case in which phase  separation  has been  neglected.
As it can be seen in this  figure the first three color  indices  show a
pronounced turn--off at low luminosities,  whereas the $B-V$ color index
does not.  In the infrared colors, intrinsically faint white dwarfs with
hydrogen--dominated atmospheres should be bluer than previously thought,
as first  discussed  by Hansen  (1998) and Saumon \&  Jacoboson  (1999).
This behavior is due to the blocking effect in the infrared of the H$_2$
collision--induced absorption.

\begin{figure}[t]
\centerline{\psfig{file=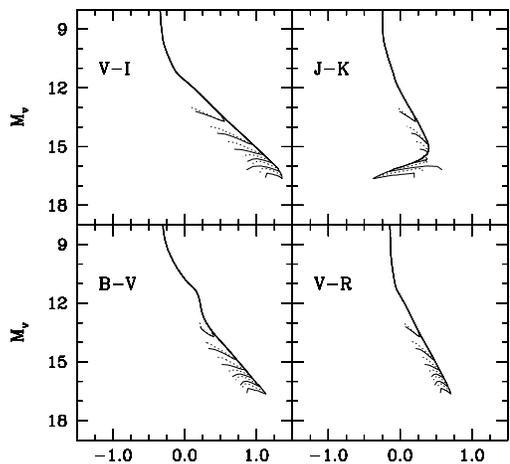,width=2.90in}}
\vskip-0.4cm
\caption{
Isochrones  for the same color--magnitude diagrams of figure
4.  See text for details.
}
\label{fig:fig5}
\end{figure}


Finally, in figure 5 we show white dwarf  isochrones for $t=2$, 4, 6, 8,
10 and 12 Gyr.  They are  particularly  useful when  studying  the white
dwarf populations in stellar  clusters, for which the progenitors can be
assumed to be coeval  and with the same  initial  chemical  composition.
Solid lines  correspond to the case in which phase  separation  has been
neglected,  whereas  the dotted  lines  correspond  to the case in which
chemical  fractionation upon  crystallization  has been fully taken into
account.  The calculation of isochrones involves two factors in addition
to  the   cooling   sequences.  First   and   most   important   is  the
initial--final  mass  relation,  second  is  the  lifetime  (up  to  the
thermally  pulsing phase) of the white dwarf  progenitors.  For both, we
have  adopted the results of Salaris et al (1997) for solar  metallicity
progenitors.  The  isochrones  were  thus  computed   self--consistently
because  the same  evolutionary  code  was  used  for  deriving  all the
required evolutionary data.

We would like to stress that, since the initial--final mass relationship
is  almost  flat in the low  mass  regime  and the  mass--main  sequence
lifetime  relationship  is very  steep, any  attempt  to derive  ages of
individual field white dwarfs from the position of low mass white dwarfs
in  the  color--magnitude   diagram  is  subject  to  potentially  large
uncertainties,  since any small error in the  determination of the white
dwarf mass  translates  into a huge relative error on its total age.  It
should  also be pointed  out that  there is a very  common  tendency  to
associate  bright white  dwarfs with young  stars,  which is  inaccurate
since they can be either bright massive white dwarfs --- and, indeed, in
this case their  total age is small --- or bright low mass white  dwarfs
with a low mass  progenitor,  which has a large main sequence  lifetime,
and therefore the reverse is true (D{\'\i}az--Pinto et al 1994, Isern et
al 1999).

As it can be seen in figure 5, in all the color--magnitude  diagrams the
isochrones show a pronounced  turn-off at their dimmer end which, as the
age of the isochrone  increases,  it is located at  increasingly  larger
magnitudes.  The presence of this  turn--off is due to the  contribution
of the most  massive  white  dwarfs,  while  the  upper  portion  of the
isochrones  closely  resembles  the  cooling  track  of  a  $\sim$  0.54
$M_{\sun}$ object.  The shape of the turn-off is modulated, in the $J-K$
colours (for ages larger than $t=7-8$ Gyr), by the intrinsic turn to the
blue of the individual cooling tracks clearly seen in figure 4.  For the
$V-I$ and $V-R$  colors the turn to the blue of the  individual  cooling
sequences begins to contribute  significantly to the isochrones for ages
$t=13-14$ Gyr.

\section{Conclusions}

In this work we have  computed  the  cooling  sequences  of very cool DA
white dwarfs with C/O cores.  These cooling  sequences  include the most
accurate physical  description of the thermodynamic  quantities for both
the core and the envelope  and have been  computed  with a well  tested,
self--consistent  evolutionary  code.  We  have  computed  as  well  the
release of latent heat using the most up to date physical inputs for the
liquid--solid  phase  transition  and we have  found  that  for the most
recent  equations of state of the degenerate  core the release of latent
heat amounts to $\sim 0.8 k_{\rm B}T$ per  particle.  Additionally,  the
release of gravitational  energy  associated to phase separation  during
crystallization  has  been  also  properly  taken  into  account.  Color
indices for several  bandpasses  have also been  computed  from the most
recent  synthetic  spectra.  Our major  findings  can be  summarized  as
follows.  Firstly, we have found that the most recent cooling  sequences
of Hansen (1999) considerably underestimate the cooling age of C/O white
dwarfs  with  hydrogen   dominated   atmospheres   even  when   chemical
fractionation  is neglected.  We have traced back the possible source of
discrepancy  and we have  found  that at a given  core  temperature  the
$T_{\rm c}-L$ relationship of Hansen (1999) differs by roughly 30\% with
respect  to other  evolutionary  calculations  computed  with  the  same
physical inputs.  This is true not only when the comparison is done with
the cooling sequences  reported here but also with the cooling sequences
by other  authors  (namely  Althaus \&  Benvenuto  1998).  The  ultimate
reason  for  this  discrepancy  remains  unidentified  although  we have
explored  several  possible  causes  without much success.  Secondly, we
have also found that a conservative  lower limit to the accumulated time
delay introduced by the release of  gravitational  energy  associated to
phase  separation is roughly 10\% at  $\log(L/L_{\sun})=  -4.5$, in good
agreement with the results of Isern et al (2000).

We  have   transformed   our  cooling   sequences   from  the   $\log(L/
L_{\sun})$-$\log(T_{\rm   eff})$  plane  into  various  color--magnitude
diagrams,  and found that  intrinsically  faint DA white  dwarfs  have a
pronounced  turn--off in the infrared  colors, and  therefore  are bluer
than  previously  thought, in good agreement  with the results of Hansen
(1998, 1999).  Cooling  isochrones  taking into account the evolutionary
time of the white dwarf  progenitors  --- suitable  for the  analysis of
white dwarfs in stellar  clusters --- have been  produced as well.  They
show  in all  colors  a  turn--off  at  the  fainter  end,  due  to  the
contribution of the more massive objects; this turn--off is modulated by
the intrinsic turn to the blue of the individual cooling tracks.

Finally we would like to stress the importance of having reliable models
of the  evolution  of white  dwarfs  and,  thus,  given the  substantial
differences  in the inferred  ages found by different  authors,  and the
complexity of the evolutionary codes needed to compute realistic cooling
sequences, more independent calculations are highly desirable.

\vskip 0.5cm

\noindent  {\em  Acknowledgements}  This work has been  supported by the
DGES  grants  PB97--0983--C03--02  and  PB97--0983--C03--03,  by the NSF
grant  AST~97--31438  and by the CIRIT.  We sincerely thank our referee,
B.  Hansen,  for  very  valuable   comments  and  criticism  which  have
considerably improved the original manuscript.  We also want to thank P.
Bergeron  for kindly  providing  us with model  atmospheres.  One of us,
EGB, also acknowledges the support received from Sun MicroSystems  under
the Academic Equipment Grant AEG--7824--990325--SP.

\end{document}